\documentclass[a4paper,11pt]{iopart}
\usepackage{graphicx}
\usepackage{cite}
\usepackage{color}
\usepackage{amsfonts}
\usepackage{srcltx}
\usepackage{array}
\begin{document}
\title{Entanglement formation under random interactions}
\author{Christoph Wick$^1$, Jaegon Um$^{1,2}$, and Haye Hinrichsen$^1$}
\address{$^1$ Universit\"at W\"urzburg, Fakult\"at f\"ur Physik und Astronomie, Am Hubland, \\ 97074 W\"urzburg, Germany}
\address{$^2$ Quantum Universe Center, Korea Institute for Advanced Study, Seoul 130-722, Korea}

\ead{cwick@physik.uni-wuerzburg.de, slung79@gmail.com, \\ \hspace{13mm}hinrichsen@physik.uni-wuerzburg.de}

\begin{abstract}
The temporal evolution of the entanglement between two qubits evolving by random interactions is studied analytically and numerically. Two different types of randomness are investigated. Firstly we analyze an ensemble of systems with randomly chosen but time-independent interaction Hamiltonians. Secondly we consider the case of a temporally fluctuating Hamiltonian, where the unitary evolution can be understood as a random walk on the $SU(4)$ group manifold. As a by-product we compute the metric tensor and its inverse as well as the Laplace-Beltrami for $SU(4)$.
\end{abstract}

\def\d{{\rm d}}
\def\0{\emptyset}
\def\ket#1{|#1\rangle}
\def\bra#1{\langle#1|}
\def\braket#1#2{\langle#1|#2\rangle}
\def\ketbra#1#2{|#1\rangle\langle #2|}
\newcommand{\vect}[1]{\mbox{\boldmath${#1}$}}

\def\alphavec{{\vect{\alpha}}}

\def\idendity{{\mbox{\boldmath$1$}}}

\def\comment#1{\color{red}[\textbf{comment: #1}]\color{black}}
\def\mark#1{\color{red}#1 \color{black}}

\newcommand{\GUE}{{\textsc{\tiny GUE}}}
\newcommand{\Haaralpha}{{\textsc{\scriptsize $\alphavec$}}}
\newcommand{\alphasub} {{\hspace*{-1pt}\textsc{\scriptsize $\alphavec$}}}
\newcommand{\Haarbeta}{{\textsc{\scriptsize $\vec\beta$}}}

\pagestyle{plain}

\section{Introduction}

If two initially separable quantum systems are exposed to random interactions they are expected to become entangled, exhibiting random quantum correlations. How do these quantum correlations arise as a function of time? To address this question we study the entanglement between two qubits subjected to random interactions as a function of time. The study of entanglement dynamics under random environments has attracted much interest recently, for instance, the emerging entanglement between coupled quantum systems through a bosonic heat bath \cite{Zell09}.
Although our system is oversimplified in comparison with these dissipative systems, we believe that our study may give the upper bound for the entanglement under the strong random interactions.    

In what follows we assume that the two-qubit system is initially prepared in a well-defined pure state. As examples we consider two different initial states, namely, a non-entangled pure state
\begin{equation}
\label{initial}
\rho(0)=|11\rangle\langle 11| 
\end{equation}
and in a fully entangled Bell state of the form
\begin{equation}
\label{BellState}
\rho(0)=|\phi\rangle\langle\phi|\,,\qquad  |\phi\rangle = \frac{1}{\sqrt{2}}\left( |00\rangle +  |11\rangle \right)\,,
\end{equation}
where $\{|00\rangle,|01\rangle,|10\rangle,|11\rangle\}$ denotes the canonical qubit configuration basis. Starting with the given initial state the system then evolves unitarily as
\begin{equation}
\label{eq:unitary_time_evolution_operator}
 \rho(t) = U(t) \rho(0) U^\dagger(t)\,,
\end{equation}
where the time evolution operator is determined by $U(0)=\mathbf{1}$ and 
\begin{equation}
i \hbar \partial_t U(t) = H(t)\,U(t)\,
\end{equation}
with a randomly chosen interaction Hamiltonian. 

Throughout this paper we consider two different types of randomness, namely
\begin{figure}
\centering\includegraphics[width=80mm]{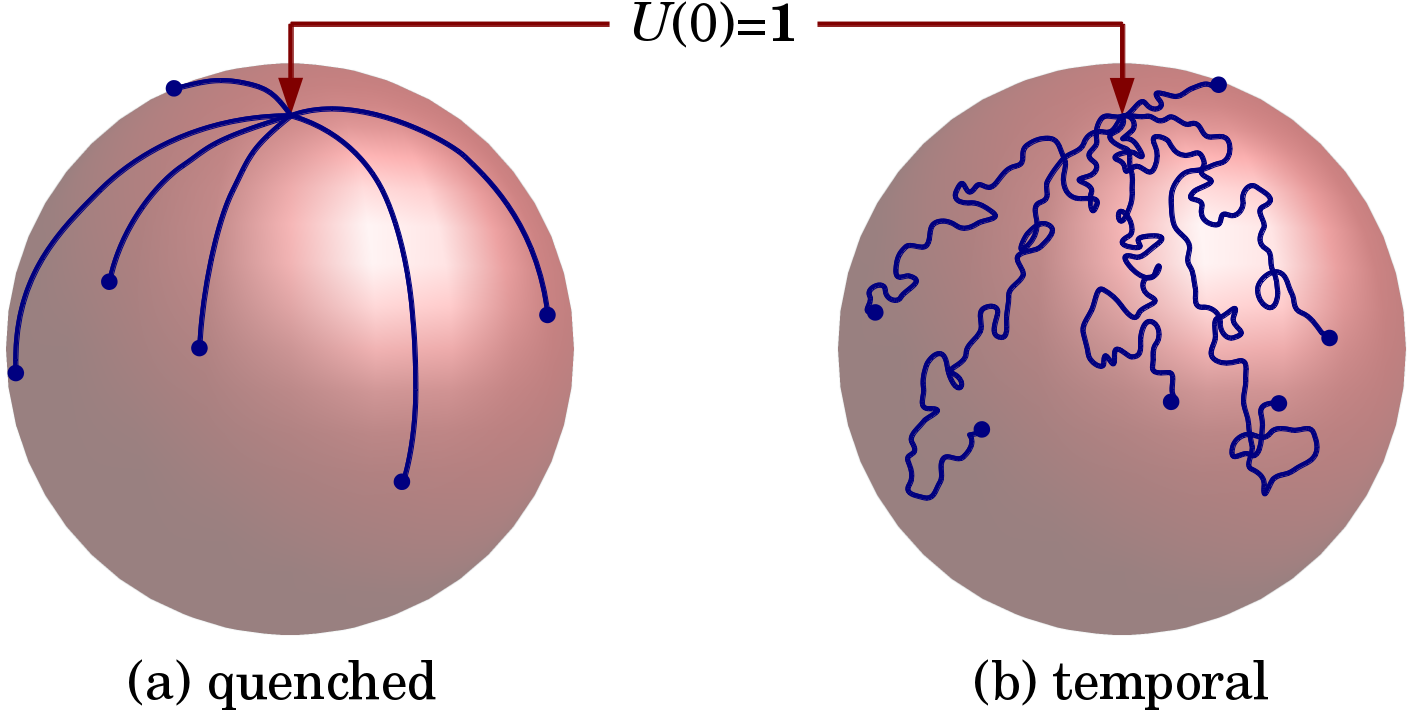}
\caption{\label{fig1}\small Over-simplified cartoon of trajectories on the $SU(4)$ group manifold, visualizing the two types of randomness discussed in the present work (see text). The north pole (red arrow) stands for the identical transformation.}
\end{figure}

\begin{itemize}
 \item[(a)]  \textbf{Quenched randomness}, where $H(t)=H$ is time-independent. In this case one considers an \textit{ensemble} of two-qubit systems starting from the same initial state, where each member evolves by a different but temporally constant Hamiltonian drawn from an $SU(4)$-invariant distribution.
 \item[(b)] \textbf{Temporal randomness}, where the dynamical evolution is generated by a time-dependent Hamiltonian $H(t)$ which fluctuates randomly as a function of time~\cite{Mehta04}. On a single system the resulting temporal evolution of the state vector can be understood as a unitary random walk in $\mathbb{C}^4$.
\end{itemize}
The difference between the two cases is visualized in Fig.~\ref{fig1}. In this figure the big red sphere stands symbolically for the 15-dimensional group manifold of $SU(4)$. Each of point on the sphere represents a certain unitary transformation $U(t)$ acting on the 4-dimensional Hilbert space of the two-qubit system. Starting with $U(0)=\mathbf 1$, which may be located e.g. at the `north pole' of the sphere, the temporal evolution $U(t)$ can be represented by a certain trajectory (blue line) on the group manifold. 

Let us now think of an \textit{ensemble} of such systems, represented by a set of statistically independent trajectories. In the quenched case (a), where a random Hamiltonian is chosen at $t=0$ and kept constant during the temporal evolution, these trajectories are straight, advancing at different pace and pointing in different directions, while in case~(b) they may be thought of as random walks on the group manifold. At a given final observation time the trajectories of the ensemble terminate in different points marked by small blue bullets in the figure, each of them representing a unitary transformation. Applying this transformation to a pure initial state one obtains a final pure state with a certain individual entanglement. In the sequel we are interested in the statistical distribution of these final states and their entanglement.  

To quantify the entanglement we use two different entanglement measures. For a pure state the entanglement is defined as the \textit{von-Neumann entropy} of the reduced states
\begin{equation}
E(t) \;=\; -\Tr\left[\rho_1(t)\ln\rho_1(t)\right] \;=\; -\Tr\left[\rho_2(t)\ln\rho_2(t)\right] \,,
\end{equation}
where $\rho_{1,2}(t)=\Tr_{2,1}\rho(t)$ denotes the time-dependent reduced density matrix of the respective qubit. In cases where the logarithm is too difficult to evaluate we resort to the so-called \textit{linear entropy}
\begin{equation}
\label{eq:linear_entropy}
L(t) =1-\Tr\left[\rho_1^2(t)\right] =1-\Tr\left[\rho_2^2(t)\right] 
\end{equation}
as an alternative entanglement measure. Note that both measures can be obtained from the more general \textit{Tsallis entanglement entropy}~\cite{Tsallis}
\begin{equation}
\label{eq:Tsallis}
E_q(t) = \frac{1-\Tr[\rho_1^q(t)]}{q-1}
\end{equation}
in the limit $q\to 1^+$ and $q \to 2$, respectively.

Furtheremore, the \textit{R\'enyi entanglement entropy}~\cite{Renyi}
\begin{equation}
\label{eq:Renyi}
H_q(t)=\frac{\log\Tr[\rho_1^q(t)]}{1-q}
\end{equation}
is of interest. Also this entropy measure generalizes the von-Neuann entropy when setting $q\rightarrow 1$.

Our main results are the following. In the first case (a), where a temporally constant Hamiltonian is randomly chosen, the mean entanglement is expected saturate at a certain value for $t \to \infty$. Our findings confirm this expectation, but surprisingly we observe that the average entanglement overshoots, i.e., it first increases, then reaches a local maximum, then slightly decreases again before it finally saturates at some stationary value. 

In the second case (b), where the Hamiltonian changes randomly as a function of time, the average entanglement saturates as well, although generally at a different level. This saturation level, which is the average entanglement of a unitarily invariant distribution of 2-qubit states, has been computed previously in Refs.~\cite{Page93,Foong94,Sen96,Kendon02}. Here we investigate the actual temporal behavior of the entanglement before it reaches this plateau. As a by-product, we compute the metric tensor and its inverse on the $SU(4)$ group manifold as well as the corresponding Laplace-Beltrami operator (see supplemental material), which to our knowledge have not been published before.

\section{Random unitary transformations in four dimensions}

\subsection{Representation of $SU(4)$ transformations}

In what follows we use a particular representation of the group $SU(4)$ which was originally introduced by Tilma \etal in Ref.~\cite{Tilma}. As reviewed in the Appendix, the group elements are generated by 15 Gell-Mann matrices $\lambda_1,\ldots,\lambda_{15}$, allowing one to parametrize unitary transformations $U\in SU(4)$ by
\begin{eqnarray} 
\label{eq:arbitrary_unitary_matrix}
U_{\Haaralpha} &=& \; e^{i\lambda_3 \alpha_1} \; e^{i\lambda_2 \alpha_2} \; e^{i\lambda_3 \alpha_3} \; e^{i\lambda_5 \alpha_4} \; \nonumber e^{i\lambda_3 \alpha_5} \; e^{i\lambda_{10} \alpha_6} \; e^{i\lambda_3 \alpha_7} \; e^{i\lambda_2 \alpha_8} \\ && \times  \; e^{i\lambda_3 \alpha_9} \; e^{i\lambda_5 \alpha_{10}} \; e^{i\lambda_3 \alpha_{11}} \; e^{i\lambda_2 \alpha_{12}} \; e^{i\lambda_3 \alpha_{13}} \; e^{i\lambda_8 \alpha_{14}} \; e^{i\lambda_{15} \alpha_{15}}\,,
\end{eqnarray}
where the 15 Euler-like angles $\alphavec=\{\alpha_1,\ldots,\alpha_{15}\}$ vary in certain ranges specified in~(\ref{alpharange}).
Applying such a unitary transformation to the non-entangled initial state $\rho(0)=|11\rangle\langle 11|$ one obtains the density matrix
\begin{equation}
\label{eq:rho_dependend_on_alpha}
\rho(\vect{\alpha}) \;=\; U_\alphasub\,\rho(0)\,U^\dagger_\alphasub
\end{equation}
with the components
\begin{eqnarray}
\label{twoqubitrepresentation}
\rho_{11}(\vect{\alpha}) &=& \cos ^2\left(\alpha _2\right) \cos ^2\left(\alpha _4\right) \sin ^2\left(\alpha _6\right)\nonumber \\
\rho_{12}(\vect{\alpha}) &=&  -\frac{1}{2} e^{2 i \alpha _1} \cos ^2\left(\alpha _4\right) \sin \left(2 \alpha _2\right) \sin
   ^2\left(\alpha _6\right) \nonumber \\
\rho_{13}(\vect{\alpha}) &=&  -\frac{1}{2} e^{i \left(\alpha _1+\alpha _3\right)} \cos \left(\alpha _2\right) \sin \left(2 \alpha _4\right) \sin ^2\left(\alpha _6\right) \nonumber \\
\rho_{14}(\vect{\alpha}) &=&  e^{i \left(\alpha
   _1+\alpha _3+\alpha _5\right)} \cos \left(\alpha _2\right) \cos \left(\alpha _4\right) \cos \left(\alpha _6\right) \sin \left(\alpha _6\right) \nonumber \\
\rho_{22}(\vect{\alpha}) &=&  \cos ^2\left(\alpha _4\right) \sin ^2\left(\alpha _2\right) \sin
   ^2\left(\alpha _6\right) \nonumber \\
\rho_{23}(\vect{\alpha}) &=&  e^{-i \left(\alpha _1-\alpha _3\right)} \cos \left(\alpha _4\right) \sin \left(\alpha _2\right) \sin \left(\alpha _4\right) \sin ^2\left(\alpha _6\right) \nonumber \\
\rho_{24}(\vect{\alpha}) &=&  -e^{-i
   \left(\alpha _1-\alpha _3-\alpha _5\right)} \cos \left(\alpha _4\right) \cos \left(\alpha _6\right) \sin \left(\alpha _2\right) \sin \left(\alpha _6\right) \nonumber \\
   \rho_{33}(\vect{\alpha}) &=&  \sin ^2\left(\alpha _4\right) \sin ^2\left(\alpha _6\right) \nonumber \\
   \rho_{34}(\vect{\alpha}) &=& -e^{i \alpha _5}
   \cos \left(\alpha _6\right) \sin \left(\alpha _4\right) \sin \left(\alpha _6\right) \nonumber \\
   \rho_{44}(\vect{\alpha}) &=& \cos ^2\left(\alpha _6\right) \,.
\end{eqnarray}
Remarkably, for $\rho(0)=|11\rangle\langle 11|$ this density matrix depends only on six angles $\alpha_1,\ldots,\alpha_6$ out of 15. Since the observables investigated in this paper are invariant under local unitary transformations, any pure separable initial state will give the same result. Therefore, without loss of generality we can restrict ourselves to $\rho(0)=|11\rangle\langle 11|$, taking advantage of the low number of parameters in this particular case.

\subsection{Computing averages on the $SU(4)$-manifold}

In the following section we will consider an ensemble of trajectories of unitary transformations generated by random interactions. Using the above representation, each trajectory can be parametrized by a time-dependent vector of Euler angles $\alphavec(t)$. A statistical ensemble of trajectories is therefore characterized by a probability density $p(\alphavec,t)$ to find a unitary transformation with the Euler angles $\alphavec$ at a given time $t$. 

For a given probability density $p(\alphavec,t)$ one can compute the ensemble average of any function $f(\alphavec)$ (such as the density matrix $\rho(\alphavec)$ or the entanglement $E(\alphavec$)) by integrating over the complete parameter space of the $SU(4)$ manifold weighted by $p(\alphavec,t)$:
\begin{equation}
\label{eq:function_average}
\Bigl\langle f(t) \Bigr\rangle_\alphasub \;=\; \frac{1}{V_{SU(4)}}\,\int_{V_{SU(4)}}  p(\alphavec,t) \, f(\alphavec) \, \d V_{SU(4)}\,.
\end{equation}
Here $V_{SU(4)}$ is the integrated group volume which serves as a normalization factor while
\begin{equation}
\label{dVsu4}
\d V_{SU(4)} \;=\; \mu(\alpha)\, \prod_{j=1}^{15} \d \alpha_j
\end{equation}
denotes the volume element on the $SU(4)$ group manifold. The actual integration measure is defined by the function $\mu(\alphavec)$ which depends on the chosen representation. Here we use the uniform measure, also known as Haar measure~\cite{Haar}, which is by itself invariant under unitary transformations.\footnote{For example, in spherical coordinates the invariant measure of the rotational group $SO(3)$ would be given by $\d V_{SO(3)}=\mu(r,\phi,\theta) \d r\,\d \theta \,\d \phi$ with $\mu(r,\phi,\theta)=r^2 \sin{\theta}$.} In the present case of $SU(4)$ with the parametrization defined above the Haar measure is defined by~\cite{Tilma}
\begin{small}\begin{equation}
\label{eq:volume_element}
\fl
\mu(\vect{\alpha})  \;=\; \sin (2 \alpha _2) \sin (\alpha _4) \sin ^5(\alpha _6) \sin (2 \alpha _8) \sin^3(\alpha _{10}) \sin (2 \alpha _{12}) \cos ^3(\alpha _4) \cos (\alpha _6) \cos (\alpha _{10}) \,. \nonumber
\end{equation}
\end{small}
The total group volume, first computed by Marinov~\cite{Marinov}, is then given by
\begin{equation}
V_{SU(4)} \;=\; \int \d V_{SU(4)} \;=\; \int\d\alpha_1\cdots\int\d\alpha_{15} \,\mu(\vect{\alpha}) \;=\;\frac{\sqrt{2}\pi^9}{3}\,.
\end{equation}
In summary, averages over the $SU(4)$ manifold weighted by the probability density $p(\alphavec,t)$ can be carried out by computing the 15-dimensional integral
\begin{equation}
\label{eq:function_average}
\Bigl\langle f(t) \Bigr\rangle_\alphasub \;=\; \frac{\sqrt{2}\pi^9}{3}\,\int\d\alpha_1\cdots\int\d\alpha_{15} \,\,\mu(\vect{\alpha})\, p(\alphavec,t) \, f(\alphavec) \,,
\end{equation}
with the measure (\ref{eq:volume_element}) and the integration ranges specified in~(\ref{alpharange}). The uniform Haar measure corresponds to taking $p(\alphavec,t)=1$.

The transformed state $\rho(\vect{\alpha})= U_\alphasub\,\rho(0)\,U^\dagger_\alphasub$ is still pure but generally entangled. Being interested in the average entanglement of states generated by random unitary transformations, it makes a difference whether the entanglement is computed before taking average over $\alphavec$ or vice versa, as will be discussed in the following.

\subsection{Average of the entanglement}

Let us first discuss the case of computing the entanglement \textit{before} taking the average over all $\alphavec$. In this case one has to compute the reduced density matrix of the first qubit for given $\rho(\alphavec)$, defined as the partial trace
\begin{equation}
\sigma(\vect{\alpha}) \;=\; \Tr_2\bigl[\rho(\vect{\alpha})\bigr]\,.
\end{equation}
For the initial state $\rho(0)=|11\rangle\langle 11|$ this is a $2\times 2$-matrix with the elements
\begin{eqnarray}
\sigma_{11}(\vect{\alpha}) &=&
\cos ^2\left(\alpha _4\right) \sin ^2\left(\alpha _6\right) \nonumber \\
\sigma_{12}(\vect{\alpha}) &=& -\frac{1}{2} e^{i \left(\alpha _1+\alpha _3\right)} \sin \left(2 \alpha _4\right) \sin ^2\left(\alpha _6\right) \cos \left(\alpha _2\right)\nonumber\\&&-\frac{1}{2} e^{-i \left(\alpha _1-\alpha _3-\alpha
   _5\right)} \sin \left(\alpha _2\right) \sin \left(2 \alpha _6\right) \cos \left(\alpha _4\right) \nonumber \\
\sigma_{22}(\vect{\alpha}) &=& \cos ^2\left(\alpha _6\right)+\sin ^2\left(\alpha _4\right) \sin ^2\left(\alpha _6\right)\,.
\end{eqnarray}
In general the reduced density matrix $\sigma(\vect{\alpha})$ is no longer pure and its von-Neumann entropy quantifies the entanglement $E(\alphavec)$ between the two qubits. In order to compute the entropy we determine the eigenvalues of $\sigma$ which are given by
\begin{eqnarray}
\fl
\kappa_{1,2}(\vect{\alpha}) \;=\;=
\frac12 \pm \frac{1}{16} \Bigl(256 \sin \left(2 \alpha _2\right) \sin \left(\alpha _4\right) \sin ^3\left(\alpha _6\right) \cos ^2\left(\alpha _4\right) \cos \left(2 \alpha_1-\alpha _5\right) \cos \left(\alpha _6\right)\nonumber
\\ \fl \hspace{38mm}
-24 \sin ^2\left(\alpha _6\right) \cos \left(2\alpha _2\right)+\cos \left(2 \alpha _6\right) \left(8-40 \sin ^2\left(\alpha _6\right) \cos \left(2 \alpha _2\right)\right)\nonumber
\\ \fl \hspace{38mm}
-32 \sin ^2\left(2 \alpha _6\right) \cos ^2\left(\alpha _2\right) \cos \left(2 \alpha _4\right) \nonumber
\\ \fl \hspace{38mm}
+32 \sin ^2\left(\alpha _2\right) \sin ^4\left(\alpha _6\right) \cos
   \left(4 \alpha _4\right)+6 \cos \left(4 \alpha _6\right)+50\Bigr)^{1/2}\,.
\end{eqnarray}
Having determined these eigenvalues, the entanglement of $\rho(\alphavec)$ is given by 
\begin{equation}
\label{entanglemententropy}
E(\vect{\alpha}) \;=\; - \sum_{i=1}^2 \kappa_i \ln \kappa_i\,.
\end{equation}
Finally, the entanglement has to be averaged over all trajectories (see Eq.~(\ref{eq:function_average})), i.e.
\begin{equation}
\overline E \;=\; \langle E(\alphavec) \rangle_\alphasub\,.
\end{equation}
However, if the average of the von-Neumann entropy is too difficult to compute, we will also use the linear entropy
\begin{equation}
\label{eq:definition_linear_entropy}
L(\vect{\alpha}) \;=\; E_2(\vect{\alpha}) \;=\; 1-\sum_{i=1}^2 \kappa_i^2
\end{equation}
as an alternative entanglement measure. 

\subsection{Entanglement of the average}

Alternatively, we may first compute the average density matrix $\rho=\langle \rho(\alphavec) \rangle_\alphasub$ and then determine the entanglement of the resulting mixed state. To this end a suitable entanglement measure is needed. An interesting quantity in this context is Wootters concurrence~\cite{Wootters} defined by
\begin{equation}
\mathcal C(\rho) = \max(0,\lambda_1-\lambda_2-\lambda_3-\lambda_4)\,,
\end{equation}
where $\lambda_1,\ldots,\lambda_4$ are the decreasingly sorted square roots of the eigenvalues  of the matrix
\begin{equation}
\Lambda=\rho(\sigma^y \otimes \sigma^y)  \rho^* (\sigma^y \otimes \sigma^y)\,.
\end{equation}
In this expression $\sigma^y$ is the Pauli matrix while $\rho^*$ denotes the complex conjugate of $\sigma$ without taking the transpose. 

From the concurrence one can easily compute the entanglement of formation of the mixed state, which is given by
\begin{equation}
E_F(\rho) = -b\ln b-(1-b)\ln (1-b)\,,
\end{equation}
where $b=\frac12+\frac12 \sqrt{1-\mathcal C(\rho)^2}$.

\section{Quenched random interactions}

In the case (a) of quenched randomness each element of the ensemble is associated with a time-independent random Hamiltonian $H$. Since the spectral decomposition
\begin{equation}
H = \sum_{j=1}^4 E_j |\phi_j\rangle\langle\phi_j|
\end{equation}
of a randomly chosen Hamiltonian is always non-degenerate, the time evolution operator can be written as
\begin{equation}
U(t)=e^{-i H t}=\sum_{j=1}^4 e^{-i E_j t} |\phi_j\rangle\langle\phi_j | \,.
\end{equation}
Hence the state of the system evolves as 
\begin{equation}
\label{eq:rho}
\rho(t) \;=\; U(t) \rho(0) U^\dagger(t) \;=\; \sum_{j,k=1}^4 e^{ -i \left(E_j - E_k\right) t }  \,\langle\phi_j|\rho(0)|\phi_k\rangle\,\,|\phi_j\rangle\langle\phi_k|\,,
\end{equation}
where $\rho(0)$ denotes the initial state. 

The Hamiltonian itself has to be drawn from a certain probabilistic ensemble of Hermitian random matrices~\cite{Mehta04,Anderson}. Here the most natural choice is again the Gaussian unitary ensemble (GUE). This ensemble has the nice property that the probability distributions for the eigenvalues $E_j$ and the eigenvectors $|\phi_j\rangle$ factorize and thus can be treated independently. More specifically, the eigenvalues are known to be distributed as
\begin{equation}
P(E_1, \dots, E_4) \;\propto\; e^{-A \sum_j E_j^2} \prod_{n>m} \left(E_n - E_m\right)^2  \,,
\end{equation}
where $A = \frac{1}{2 \sigma^2}$ is a constant determining the width of the energy fluctuations and therewith the time scale of the temporal evolution. In the following the corresponding average over the energies will be denoted by $\langle\ldots\rangle_E$. On the other hand, the orthonormal set of eigenvectors is randomly oriented in the four-dimensional Hilbert space according to Haar measure, independent of the eigenvalues. If one defines the qubit basis
\begin{equation}
\label{eq:qubit_basis}
\{\ket 1,\ket 2, \ket 3, \ket 4\} := \{ \ket{00},\ket{01},\ket{10},\ket{11}\}
\end{equation}
this average can be carried out by setting
\begin{equation}
\label{eq:rewrite_ev_H_as_U_i}
\ket{\phi_j} \;:=\; U_\alphasub^\dagger \ket j
\end{equation}
and integrating over all Euler angles $\vect{\alpha}$ according to the Haar measure (see Appendix~B). This average will be denoted by $\langle\ldots\rangle_\alphasub$. The total GUE average thus factorizes as
\begin{equation}
\langle\ldots\rangle_\GUE = \langle\ldots\rangle_E \langle\ldots\rangle_\alphasub.
\end{equation}

\subsection{Entanglement of the averaged density matrix}
Let us now compute the average density matrix
\begin{equation}
\label{rhoav}
\Bigl\langle\rho(t)\Bigr\rangle_\GUE \;=\; \sum_{j,k=1}^4 \underbrace{\Bigl\langle e^{ -i \left(E_j - E_k\right) t } \Bigr\rangle_E}_{R_{jk}}\,\,\underbrace{\Bigl\langle|\phi_j\rangle\langle\phi_j|\,\rho(0)\,|\phi_k\rangle\langle\phi_k|\Bigr\rangle_\Haaralpha}_{\mathbf{T}_{jk}}\,.
\end{equation}
First we compute the average over the energies
\begin{equation}
R_{jk}  \;=\; \frac{1}{\mathcal N}\int_{-\infty}^{+\infty}\d E_1 \cdots \d E_4 \, P_\GUE(E_1, \dots, E_4) \, e^{ -i \left(E_j - E_k\right) t } \,,
\end{equation}
where $\mathcal N={  \int_{-\infty}^{+\infty} \d E_1 \cdots \d E_4 \, P_\GUE(E_1, \dots, E_4) }=\frac{9\pi}{2A^8}$ is the normalization factor. This leads us to the result

\begin{equation}
R_{jk} \;=\; 
f(\tau)+\Bigl(1-f(\tau)\Bigr)\delta_{jk} \;=\;
\left\{ \begin{array}{cc} 1 & \mbox{ for } j = k \\ 
f(\tau) & \mbox{ for } j \neq k  \end{array} \right. \,,
\end{equation}
where we defined the scaled time 
\begin{equation}
\tau := t/\sqrt{2A} 
\end{equation}
and the function
\begin{equation}
\label{goftau}
f(\tau) = \frac{1}{72} e^{-\tau ^2} \left(-2 \tau ^{10}+25 \tau ^8-128 \tau ^6+276 \tau ^4-288 \tau ^2+72\right)\,.
\end{equation}
Thus Eq.~(\ref{rhoav}) reduces to
\begin{equation}
\label{splitaverage}
\Bigl\langle\rho(t)\Bigr\rangle_\GUE \;=\; f(\tau)\sum_{j,k=1}^4 \mathbf{T}_{jk} +\Bigl(1-f(\tau)\Bigr) \sum_{j=1}^4 \mathbf{T}_{jj}
\end{equation}
What remains is to determine the operators
\begin{equation}
\mathbf{T}_{jk} \;=\; \Bigl\langle \ketbra{\phi_j}{\phi_j}\; \rho(0) \;\ketbra{\phi_k}{\phi_k}\Bigr\rangle_\alphasub = \Bigl\langle U^\dagger_\alphasub\ketbra{j}{j}U_\alphasub\; \rho(0) \;U^\dagger_\alphasub \ketbra{k}{k}U_\alphasub \Bigr\rangle_\alphasub.
\end{equation}
Obviously, the first sum in Eq.~(\ref{splitaverage}) is given by
\begin{equation}
\label{eq:sum_Tij_equals_rho0}
\sum_{j,k=1}^4 \mathbf{T}_{jk} \;=\;  \Bigl\langle U^\dagger_\alphasub U_\alphasub\; \rho(0) \;U^\dagger_\alphasub U_\alphasub \Bigr\rangle_\alphasub \;=\; 
\Bigl\langle \rho(0)\Bigr\rangle_\alphasub \;=\; \rho(0).
\end{equation}
As for the second sum in Eq.~(\ref{splitaverage}), we note that the distribution of eigenvectors $\ket{\phi_j} \;=\; U_\alphasub^\dagger \ket j$ is invariant under a permutation of the basis vectors $\ket j$, hence the four operators $\mathbf{T}_{jj}$ coincide. Moreover, under a unitary transformation $V \in SU(4)$ they transform as
\begin{eqnarray}
V \mathbf{T}_{jj} V^\dagger &=& \Bigl\langle V U^\dagger_\alphasub\ketbra{j}{j}U_\alphasub\; \rho(0) \;U^\dagger_\alphasub \ketbra{j}{j}U_\alphasub V^\dagger \Bigr\rangle_\alphasub \\
&=& \Bigl\langle U^\dagger_\alphasub\ketbra{j}{j}U_\alphasub\; \Bigl(V\rho(0)V^\dagger\Bigr) \;U^\dagger_\alphasub \ketbra{j}{j}U_\alphasub  \Bigr\rangle_\alphasub\,,\nonumber
\end{eqnarray}
where we have used the invariance of the GUE-eigenvectors under the replacement $U_\alphasub \to U_\alphasub V$. This means that $\mathbf{T}_{jj}$ is invariant under $V$ if and only if $V$ commutes with the initial state $\rho(0)$. For a pure initial state this implies that $\mathbf{T}_{jj}$ has to be a linear combination of the identity and the initial state itself, i.e. $\mathbf{T}_{jj}=a \mathbf{1}+b\rho(0)$. The linear coefficients $a$ and $b$ can be determined as follows. On the one hand, the identity
\begin{equation}
1 \;=\; \Tr\Bigl[\langle\rho(t)\rangle_\GUE \Bigr] \stackrel{(\ref{splitaverage},\ref{eq:sum_Tij_equals_rho0})}{=} f(\tau)+(1-f(\tau) ) \Tr\Bigl[\sum_{j=1}^4 \mathbf{T}_{jj} \Bigr]
\end{equation}
implies that $\Tr\Bigl[\sum_{j=1}^4 \mathbf{T}_{jj} \Bigr] = 1$, hence $4a+b=1/4$. On the other hand, we note that
\begin{eqnarray}
\Tr\Bigl[\rho(0) \mathbf{T}_{jj} \Bigr] &=&
\Bigl\langle  \Tr\Bigl[ \rho(0) \ketbra{\phi_j}{\phi_j} \rho(0) \ketbra{\phi_j}{\phi_j} \Bigr] \Bigr\rangle_\alphasub 
\;=\;\Bigl\langle  \bra{\phi_j} \rho(0) \ket{\phi_j}^2 \Bigr\rangle_\alphasub 
\end{eqnarray}
is invariant under unitary transformations of $\rho(0)$ and independent of $j$, hence we may choose $j=4$ and $\rho(0)=\ketbra44$ to obtain
\begin{equation}
\Tr\Bigl[\rho(0) \mathbf{T}_{jj} \Bigr] \;=\; \Bigl\langle  \braket{\phi_4}4^2 \braket4{\phi_4}^2 \Bigr\rangle_\alphasub 
\;=\; \Bigl\langle  \cos^4 (\alpha_6) \Bigr\rangle_\alphasub = \frac1{10}\,,
\end{equation}
giving $a=b=\frac{1}{20}$. Therefore, we arrive at the convex combination of ${\bf 1}/4$ and $\rho(0)$
\begin{equation}
\label{averagedensitymatrix}
\bar{\rho}(t) =
\Bigl\langle\rho(t)\Bigr\rangle_\GUE \;=\; \frac{1- f(\tau)}{5} {\bf 1} +\frac{1+4 f(\tau)}{5} \rho(0) 
\end{equation}
with $f(\tau)$ given in Eq. (\ref{goftau}) and $\tau = t/\sqrt{2A}$, which holds for any pure initial state $\rho(0)$.
As expected, the averaged state lies on the segment between the initial state $\rho(0)$ and the maximally mixed state ${\bf 1} / 4$ due to the symmetries of the Haar measure of $SU(4)$ .

\begin{figure}
\centering\includegraphics[width=150mm]{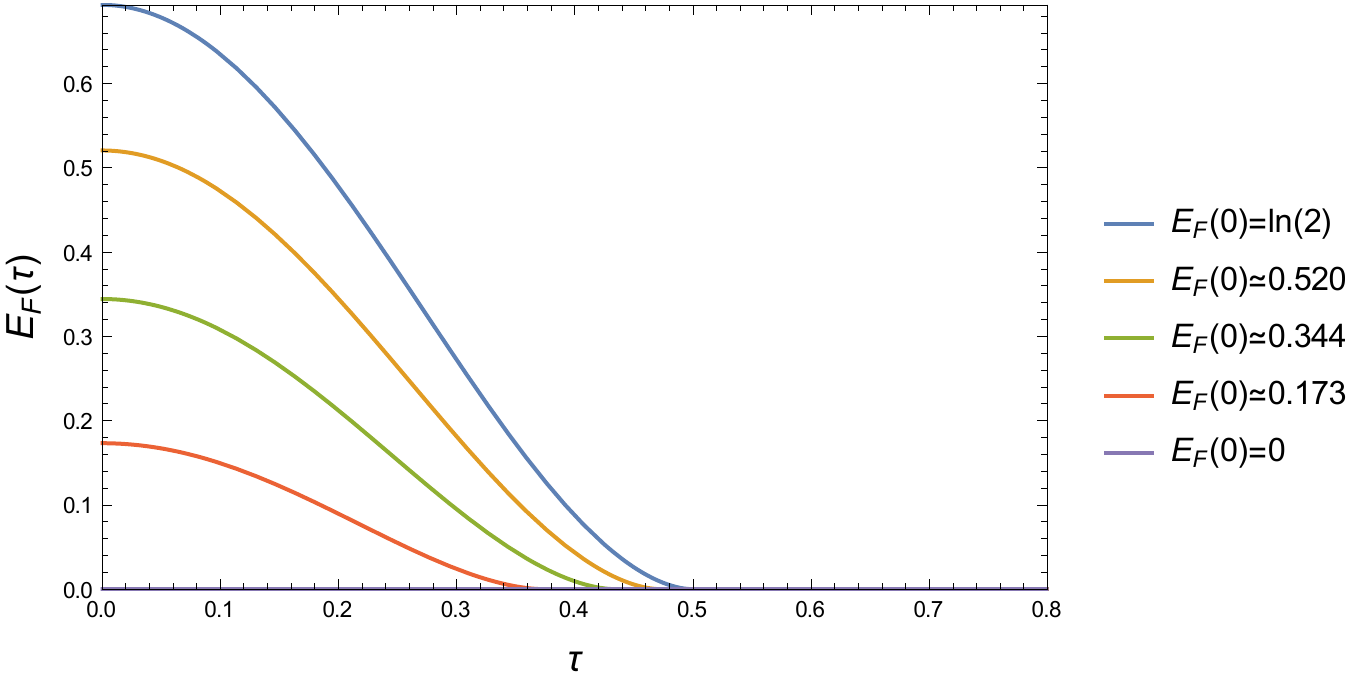}
\caption{\label{fig2}\small Quenched case: Analytically calculated entanglement of formation of the averaged density matrix in Eq.~(\ref{averagedensitymatrix}) as a function of the scaled time $\tau=t/\sqrt{2A}$ using different initial conditions $\rho(0)=\ket\psi\bra\psi$ of pure states $\ket{\psi_c}=(c,0,0,\sqrt{1-c^2})$ with $c=\{0,0.204,0.330,0.464,1/\sqrt 2\}$ from bottom to top. In particular, the purple and the blue line represent a non-entangled and maximally entangled initial state, respectively. }
\end{figure}

Having computed the mixed state of the ensemble $\bar \rho(t)$ we can now compute the corresponding entanglement of formation as a function of time. For a non-entangled initial pure state $\rho(0)=|11\rangle\langle 11|$ we find that $E_F(\bar\rho(t))=0$ for all times.
However, if we start from the Bell state (\ref{BellState}) with the initial entanglement $E_F(\rho(0))=\ln 2\approx 0.693$ we find numerically that the entanglement first decreases and vanishes at the finite scaled time $\tau\approx 0.4997.$ (see Fig.~\ref{fig2}).

\subsection{Average of the individual entanglement}
Instead of computing the entanglement of the average density matrix, let us now compute the average of the individual entanglement of each trajectory, i.e. the entanglement is computed before taking the GUE average. Although the von-Neumann entanglement entropy of the individual pure states would be straight-forward to compute (see (\ref{entanglemententropy})), we did not succeed to compute the average. For this reason let us consider the GUE average of the linear entropy 
\begin{equation}
\Bigl\langle L(t) \Bigr\rangle_\GUE \;=\; 1- \Bigl\langle \rho_1^2(t) \Bigr\rangle_\GUE
\end{equation}
where $\rho_1(t)$ denotes the reduced density matrix of the first qubit. In the qubit basis~(\ref{eq:qubit_basis}) this can be rewritten as
\begin{equation}
\Bigl\langle L(t) \Bigr\rangle_\GUE \;=\; 1- \sum_{\mu,\beta,\gamma,\delta=1}^2 \Bigl\langle \,\langle \mu\beta|\rho(t)|\gamma\beta\rangle\,\langle \gamma\delta |\rho(t) |\delta\mu\rangle\,\Bigr\rangle_\GUE
\end{equation}
Inserting (\ref{eq:rho}) and exploiting again that the GUE average factorizes, we get
\begin{eqnarray}
\Bigl\langle L(t) \Bigr\rangle_\GUE &=& 1-
\sum_{\mu,\beta,\gamma,\delta=1}^2 \sum_{j,k,l,m=1}^4 
\Bigl\langle e^{i(E_j-E_k+E_l-E_m)t}\Bigr\rangle_E \times \\
&& \hspace{10mm}
\Bigl\langle  {c^{\mu\beta}_{j}}^* \,\langle\phi_j|\rho(0)|\phi_k\rangle \,c^{\gamma\beta}_{k} \,
{c^{\gamma\beta}_{l}}^* \, \langle \phi_l| \rho(0)| \phi_m\rangle\, c^{\delta\mu}_{m}  \Bigr\rangle_\alphasub \,,\nonumber
\end{eqnarray}
where $c^{\mu\beta}_j=\langle\phi_j | \mu\beta\rangle$. For the initially non-entangled state $\rho(0) = \ketbra{11}{11}$ this expression reduces further to
\begin{eqnarray}
\Bigl\langle L(t) \Bigr\rangle_\GUE  &=& 1 - \hspace{-2mm}
\sum_{\mu,\beta,\gamma,\delta=1}^2 \sum_{j,k,l,m=1}^4 
\underbrace{\Bigl\langle e^{-i (E_j - E_k + E_l - E_m) \tau \sqrt{2A}} \Bigr\rangle_E}_{R_{ijkl}(\tau)}
\\ && \hspace{30mm} \times  \nonumber
\underbrace{\Bigl\langle 
{c^{\mu\beta}_{j}}^* \, c^{11}_j \, {c^{11}_k}^*  \,c^{\gamma\beta}_{k} \,
{c^{\gamma\beta}_{l}}^* \, c^{11}_l \, {c^{11}_m}^*  \, c^{\delta\mu}_{m} 
\Bigr\rangle_\alphasub}_{\mathbf{T}_{jklm}^{\mu\beta\gamma\delta}}\,.
\end{eqnarray}
with the scaled time $\tau =t/\sqrt{2 A}$. As shown in \ref{appendix:integrate_R_and_T}, the averages $ R_{ijkl}(\tau)$ $\mathbf{T}_{jklm}^{\mu\beta\gamma\delta}$ can be computed directly by integration over the given probability distributions in GUE, leading us to the final result
\begin{eqnarray}
\label{eg:quenched_av_lin_entropy}
\langle L(\tau)\rangle_\GUE =
-\frac{1}{630} e^{-2 \tau ^2} \left(32 \tau ^8-128 \tau ^6+168 \tau ^4-72 \tau ^2+9\right) \\
-\frac{1}{840} e^{-\tau ^2} \left(-2 \tau ^{10}+25 \tau ^8-128 \tau ^6+276 \tau ^4-288 \tau ^2+72\right) \nonumber \\
-\frac{1}{420} e^{-3 \tau ^2} \left(-54 \tau ^{10}+387 \tau ^8-832 \tau ^6+828 \tau ^4-288 \tau ^2+24\right) \nonumber \\
-\frac{1}{315} e^{-4 \tau ^2} \left(-256 \tau ^{10}+800 \tau ^8-1024 \tau ^6+552 \tau ^4-144 \tau ^2+9\right)+\frac{13}{70}\,.\nonumber
\end{eqnarray}

\begin{figure}
\centering\includegraphics[width=150mm]{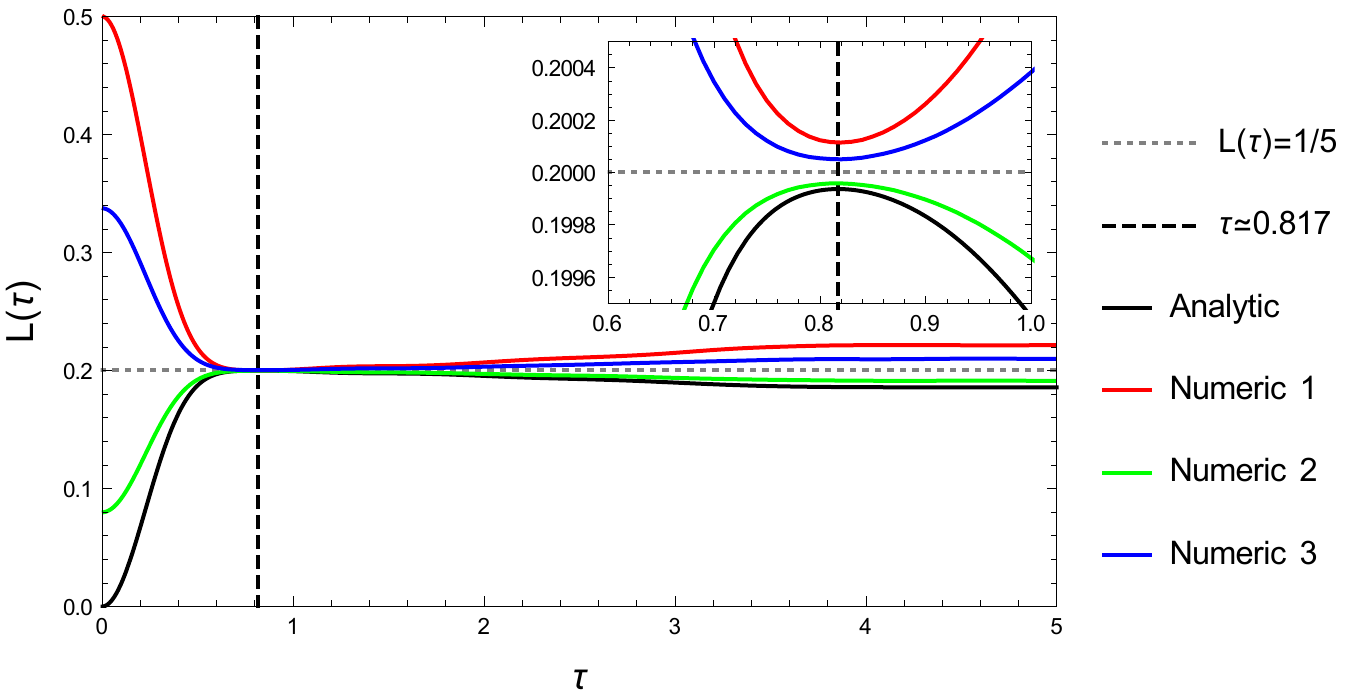}
\caption{\label{fig3}\small Average linear entanglement entropy according to Eq.~(\ref{eg:quenched_av_lin_entropy}) as a function of the scaled time $\tau=t/\sqrt{2A}$ with $c=\{0,0.204,0.464,1/\sqrt 2\}$ from bottom to top. The numerical data (see. \ref{sec:numeric_calculations}) has computation errors smaller than the thickness of the lines. The upper right panel shows the magnified area around the touching point at $\tau_{max}$ (marked by the dashed line).}
\end{figure}

\noindent
This function is plotted in Fig.~\ref{fig3}. As one can see, the linear entanglement entropy (black line) first increases rapidly, then reaches a local maximum $\langle L\rangle \simeq 0.199936<0.2$ at $\tau_{max} \simeq 0.817377$, then decreases again and finally saturates at the value
\begin{equation}
\lim_{\tau\rightarrow\infty}\langle L(\tau)\rangle_\GUE = 13/70 \simeq 0.1857\,.
\end{equation}
Because it would need much more effort to calculate the analytical the linear entropy for different initial states analytically, we used numerical methods. The results are compared in Fig.~\ref{fig3}. As one can see clearly, all the lines tend to touch the limit value at the fixed time $\tau\simeq 0.817$ and the curves do not intersect.

Fig.~\ref{fig1} explains the meaning of this result: Each single trajectory of the ensemble on the surface is deterministic with a given initial point, direction and velocity. Since all members of the ensemble share the same initial starting point on the \textit{upper half} of the sphere, the probability for finding the walkers can be slightly higher on the upper half in the long-time limit because all trajectories will periodically return to this point. This is why the limit depends on the initial state and therefore deviates from the Haar measure. The bump can be seen as a transient state in which the probability distribution seems to be almost randomly distributed before saturating.

Note that in contrast to the case discussed before (see Fig.~\ref{fig2}) the system remembers its initial state, saturating at different levels of entanglement in the limit $t \to \infty$.

\section{Time-dependent random interactions}

Let us now consider the case (b) of a temporally varying Hamiltonian, where the state vectors of the ensemble perform a unitary random walk on the $SU(4)$ manifold. In this case the quantity of interest is the probability distribution $p(\alphavec,t)$ to find the time evolution operator $U(t)$ with the Euler angles $\alphavec$ at time $t$. This probability distribution allows one to compute the ensemble average of any function $f(\alphavec)$ (such as the density matrix $\rho(\alphavec)$ or the entanglement $E_\alphavec$) by integration over the complete $SU(4)$ volume weighted by $p(\alphavec,t)$, i.e. we have to compute the integral
\begin{equation}
\label{eq:function_average}
\Bigl\langle f(t) \Bigr\rangle = \frac{1}{V_{SU(4)}}\, \int_{V_{SU(4)}}  f(\alphavec) \, p(\alphavec,t) \, \d V_{SU(4)}
\end{equation}
over the ranges specified in~(\ref{alpharange}), where $\d V_{SU(4)}$ denotes the volume element according to the Haar measure defined in Eq. (\ref{dVsu4}).

\subsection{Expected average entanglement of a uniform distribution}

Before studying the temporal evolution in detail, let us consider the limit $t \to \infty$, where we expect the state vectors to be uniformly distributed on the group manifold. Since such an ensemble is by itself invariant under unitary transformations, the state vectors are distributed according to a Gaussian Unitary Ensemble (GUE). Starting from this observation, Page~\cite{Page93} conjectured a closed expression for the expected average entanglement of an arbitrary bipartite quantum system with Hilbert space dimensions $m$ and $n$, which was later proven rigorously by Foong, Kanno, and Sen~\cite{Foong94,Sen96}. This formula describes the average entanglement of a random pure state between two subsystems with Hilbert space dimensions $m$ and $n$:
\begin{equation}
\Bigl\langle E_{m,n} \Bigr\rangle_\GUE \;=\;  \left( \Bigl( \sum_{k=n+1}^{mn} \frac{1}{k}\Bigr) - \frac{m - 1}{2n} \right)\,.
\end{equation}
Applying this formula to a random two-qubit system, one obtains
\begin{equation}
\bigl\langle E \bigr\rangle_\GUE \;=\; \bigl\langle E_{2,2} \bigr\rangle \;=\; \frac{1}{3}\,.
\end{equation}
This is the average entanglement between two qubits in a randomly chosen pure state.

\subsection{Heat conduction equation on the $SU(4)$ manifold}

In order to compute the probability distribution $p(\vect{\alphavec},t)$ one has to solve the heat conduction equation
\begin{equation}
\label{eq:heat_equation}
\frac{\partial p(\alphavec,t)}{\partial t} - D \Delta_\alphasub p(\alphavec,t) = 0
\end{equation}
on the curved $SU(4)$ group manifold. Here $D$ denotes the diffusion constant while $\Delta_\alphasub$ is the so-called Laplace-Beltrami operator which generalizes the ordinary Laplacian on a curved space. As stated by \cite{Nechita13} this Laplace-Beltrami operator is the Markov generator of the unitary brownian motion (see \ref{appendix:derive_diffusuion_equation}). On a Riemannian manifold with the metric tensor~$g_{ij}$ the Laplace-Beltrami-Operator $\Delta$ is given by
\begin{equation}
\label{eq:definition_laplace_beltrami}
\Delta f = \frac{1}{\sqrt{|g|}} \partial_i \left( \sqrt{|g|} g^{ij} \partial_j f\right)\,,
\end{equation}
where $g^{ij}$ are the components of the inverse metric tensor and $\sqrt{|g|}=\sqrt{\det g}$.

To our best knowledge the explicit expressions for $g_{ij}$, $g^{ij}$ and $\Delta_\alphasub$ have not been published before. This is perhaps due to the fact that the formulas are so complex that even powerful computer algebra systems such as Mathematica$^{\mbox{\tiny\textregistered}}$ are not able to compute the inverse metric directly. Instead one has to invert the matrix manually element by element. Our explicit results are included in the supplemental material attached to this paper.

\subsection{Early-time expansion}

The solution $p(\alphavec,t)$ of the heat conduction equation (\ref{eq:heat_equation}) and as well the averaged function $\langle f(t) \rangle$ can be expanded as a Taylor series around $t=0$:
\begin{equation}
\label{eq:taylor_propability}
p(\alphavec,t) \;=\; \sum_{n=0}^{\infty} \frac{t^n}{n!} \left. \frac{\partial^n}{\partial t^n}p(\alphavec, t)\right|_{t=0}
\\
\label{eq:taylor_function}
\end{equation}
\begin{equation}
\Bigl\langle f(t) \Bigr\rangle \;=\; \sum_{n=0}^{\infty} \frac{t^n}{n!} \left. \frac{\partial^n}{\partial t^n} \Bigl\langle f(t) \Bigr\rangle \right|_{t=0}
\end{equation}
Using (\ref{eq:function_average}) we can compare the coefficients of the both Taylor series. giving
\begin{equation}
\frac{\partial^n}{\partial t^n} \Bigl\langle f(t) \Bigr\rangle \Big|_{t=0}
\;=\; \frac{1}{V_{SU(4)}}\, \int \d V_{SU(4)}\,\, f(\alphavec) \,  \frac{\partial^n}{\partial t^n} p(\alphavec,t) \Big|_{t=0} \,,
\end{equation}
where $\d V_{SU(4)}=\sqrt{|g|}\d^{15}\alpha$ denotes the volume element defined in Eq. (\ref{dVsu4}). Using the heat equation  (\ref{eq:heat_equation}) we can replace the partial derivative, obtaining
\begin{equation}
\label{eq:equating_coefficients_intermediate}
\frac{\partial^n}{\partial t^n} \Bigl\langle f(t) \Bigr\rangle \Bigg|_{t=0}
\;=\;  \frac{1}{V_{SU(4)}}\, D^n\int \d V_{SU(4)}\,\,  f(\alphavec)\, \Delta^n_\alphasub p(\alphavec,t) \Big|_{t=0}\,.
\end{equation}
The r.h.s. is an integral over derivatives of the probability density $p(\alphavec,t)$ evaluated at $t=0$. If all trajectories start at $\alphavec_0$ it is easy to see that this probability density at $t=0$ is given by 
\begin{equation}
\label{eq:boundary_condition}
p(\alphavec,t=0) \;=\; \frac{V_{SU(4)}}{\sqrt{|g|}} \delta(\alphavec - \alphavec_0).
\end{equation}
Inserting this expression into (\ref{eq:equating_coefficients_intermediate}) the integral can be evaluated by partial integration, giving
\begin{equation}
\label{eq:equating_coefficients}
\frac{\partial^n}{\partial t^n} \langle f(t) \rangle \Bigg|_{t=0}
= D^n \Delta^n_\alphasub f(\alphavec) \Bigg|_{\alpha=\alpha_0}
\end{equation}

\subsection{Average of the density matrix}
\label{AverageDensSection}
Using (\ref{eq:rho_dependend_on_alpha}) we calculate the derivatives $\triangle^n_\alphasub \rho(\vect{\alpha})$.
We find that 
\begin{equation}
\triangle^1_\alphasub \rho(\vect{\alpha}) = -8 \, \rho(\alphavec) + 2 \cdot \idendity.
\end{equation} 
Therefore we can express all higher derivatives of $n\geq 1$ in terms of the first derivative\footnote{Although this remarkable property calls for a deeper reason, we have no convincing explanation so far.}
\begin{equation}
\frac{\partial^n}{\partial t^n} \Bigl\langle \rho(t) \Bigr\rangle \Bigg|_{t=0}= D^n (-8)^{n-1} \triangle_\alphasub \rho(\vect{\alpha})\,.
\end{equation}
Hence, the solution of the averaged density matrix can be written as
\begin{equation}
\label{eq:tr_av_rho_general}
\Bigl\langle \rho(t) \Bigl\rangle = \frac{1}{4} \cdot \idendity + \left(\rho(\vect{\alpha}_0) - \frac{1}{4} \cdot \idendity \right) e^{-8 D t}  
\end{equation}
Using the non-entangled initial state $\ket{\psi_0} = \ket{11}$ (which corresponds to taking $\vect{\alpha}_0=0$) this result specializes to
\begin{equation}
\label{eq:tr_av_rho_unent_initial_state}
\Bigl\langle \rho(t) \Bigr\rangle_{\alpha_0 = 0} = \left(
\begin{array}{cccc}
 \frac{1}{4}-\frac{1}{4} e^{-8 D t} & 0 & 0 & 0 \\
 0 & \frac{1}{4}-\frac{1}{4} e^{-8 D t} & 0 & 0 \\
 0 & 0 & \frac{1}{4}-\frac{1}{4} e^{-8 D t} & 0 \\
 0 & 0 & 0 & \frac{1}{4}+\frac{3}{4} e^{-8 D t} \\
\end{array}
\right)\,.
\end{equation}
As can be seen, this density matrix relaxes exponentially and becomes fully mixed in the limite $t \to \infty$. 

\subsection{Average of the linear entropy}

The same calculation can be carried out for the linear entropy defined in  (\ref{eq:definition_linear_entropy}).
Here we find that
\begin{equation}
\triangle^1_\alphasub L(\alphavec) = -20 \, L(\alphavec) + 4.
\end{equation}
For this reason, the calculation is completely analogous, giving
\begin{equation}
\label{eq:tr_av_L_general}
\Bigl\langle L(t) \Bigr\rangle = \frac{1}{5} + \left( L(\alphavec_0) - \frac{1}{5} \right) e^{-20 D t}.
\end{equation}
The result for an non-entangled initial state $L(\alphavec_0) = 0$ is therefore
\begin{equation}
\label{eq:tr_av_L_unent_initial_state}
\Bigl\langle L(t) \Bigr\rangle = \frac{1}{5} - \frac{1}{5} e^{-20 D t}.
\end{equation}
For a fully entangled initial state $L(\alphavec_0) = 1/2$ we get
\begin{equation}
\label{eq:tr_av_L_fullent_initial_state}
\Bigl\langle L(t) \Bigr\rangle = \frac{1}{5} + \frac{3}{10} e^{-20 D t}.
\end{equation}
In both cases the averaged linear entropy tends towards $\frac{1}{5}$ which coincides with the value of a randomly distributed density matrix according to Haar-measure (cf. Ref. \cite{Lubkin77,Page93}).

\subsection{Average of the Tsallis entropy}

Because of computational difficulties we calculated the Tasllis and von-Neumann entropy only calculated to first order in $t$. Moreover, we restricted ourselves to the case of a fully entangled initial state since the Taylor expansion fails for a non-entangled initial state due to a logarithmic factor in time at $t=0$.

Using the definition of the Tsallis entropy (\ref{eq:Tsallis}) and following the same lines as outlined above, one ends up with the first order approximation
\begin{equation}
\label{eq:tr_approx_av_Tasllis_fullent_init_state}
\Bigl\langle E_q(t) \Bigr\rangle = \frac{1-2^{1-q}}{q-1} -3 \cdot 2^{2-q} q D t + \mathcal{O}(t^2).
\end{equation}
In the limit $q\rightarrow 2$ this reproduces the first order terms of (\ref{eq:tr_av_L_fullent_initial_state}) whereas for $q\rightarrow 1$ we obtain the first-order approximation of the von-Neumann entropy
\begin{equation}
\label{eq:tr_approx_av_vanNeumann_fullent_init_state}
\Bigl\langle E(t) \Bigr\rangle = \log{2} - 6 D t + \mathcal{O}(t^2).
\end{equation}	

\begin{figure}
\centering\includegraphics[width=150mm]{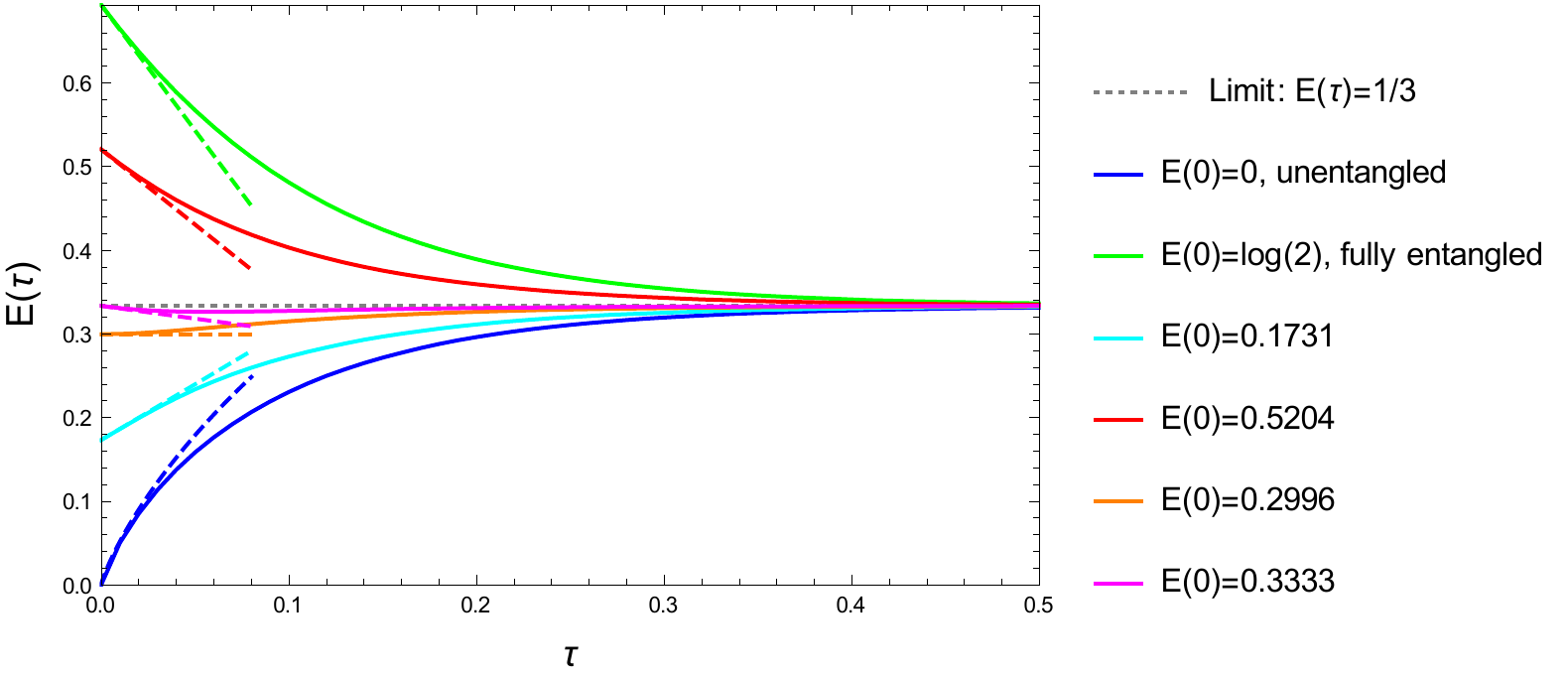}
\caption{\label{fig4}\small Numerical results (see \ref{sec:numeric_calculations}) of the von-Neumann entropy compared to the analytical approximations (dashed), starting from pure initial states $\rho(0) = \ket{\psi_c}\bra{\psi_c}$ with $c=\{0,0.203,0.300,0.322,0.464,1/\sqrt 2\}$ (see Fig.~\ref{fig2}). The error of the numerical calculation is smaller than the thickness of the lines. The scaled time is defined as $\tau = 2 D t$.}
\end{figure}

\noindent
By direct approximation on can derive the first order in $\tau = 2 D t$ of the von-Neumann entropy directly for a few timesteps using an initial state $\ket{\psi_0} = \left(\cos \phi, 0, 0,\sin \phi \right)$ with  $\phi\in(0,\pi/4)$:
\begin{eqnarray}
\label{eq:tr_approx_av_vanNeumann_abitrary_init_state}
\Bigl\langle E(\tau) \Bigr\rangle =& \left( -\log (\sin (\phi ))-\log (\cos (\phi ))-\cos (2 \phi ) \log (\cot (\phi )) \right) +\\ & \left( 2 \cos (4 \phi ) \sec (2 \phi ) \log (\cot (\phi ))-1 \right)\cdot \tau + \mathcal{O}(\tau^2) \nonumber
\end{eqnarray}
For a fully entangled initial state ($\phi\rightarrow\pi/4$) this reproduces Eq.
(\ref{eq:tr_approx_av_vanNeumann_fullent_init_state}).
Unfortunately, the expansion does not converge for $\phi\rightarrow 0$, which is why one has approximate an unentangled initial state directly, obtaining
\begin{equation}
\label{eq:tr_approx_av_vanNeumann_unent_init_state}
\Bigl\langle E(\tau) \Bigr\rangle = \tau (\gamma - \log{\tau}) + \mathcal{O}(\tau^2)\, ,
\end{equation}
where $\gamma$ is Euler’s constant with numerical value of $\gamma\approx 0.577$. 

To clarify these results Fig.~\ref{fig4} shows numerical calulations and the analytical first-order approximations as dashed lines for several initial states.

It is interesting to note that the von-Neumann entropy shows a bump if one starts near the limit value of $1/3$, whereas the linear entropy does not have such a behaviour.
But since one can interpret the linear entropy as lowest-order expansion term of the von-Neumann entropy around a pure state, it is not surprising that the von-Neumann entropy shows a more complex behavior. Moreover, a fixed initial state with a von-Neumann entropy of $1/3$ is obviously not yet randomly distributed in $SU(4)$.

\subsection{Average of the R\'enyi entropy}

We perform similar calculations as above for the R\'enyi entropy as defined in (\ref{eq:Renyi}). This leads to the following approximation in first order for small times using a fully entangled initial state:
\begin{equation}
\label{eq:tr_approx_av_Renyi_fullent_init_state}
\Bigl\langle H_q(t) \Bigr\rangle = \log 2 - 6 D q t + \mathcal{O}(t^2)
\end{equation}
For $q\rightarrow 1$ we obtain the von-Neumann entropy as already 	computed in Eq. (\ref{eq:tr_approx_av_vanNeumann_fullent_init_state}).

Setting $q=2$ one can compute the first order approxiation first order in $\tau = 2 D t$ of the R\'enyi entropy for the state $\ket{\psi_0} = \left(\cos \phi, 0, 0,\sin \phi \right)$ as defined above:
\begin{eqnarray}
\Bigl\langle H_2(\tau) \Bigr\rangle = &\log 4 -\log (\cos (4 \phi )+3) + \\ &\frac{(28 \cos (4 \phi )+3 \cos (8 \phi )+1)}{(\cos (4 \phi )+3)^2} \cdot \tau + \mathcal{O}(\tau^2) \nonumber
\end{eqnarray}
Note that this equation holds for all $\phi \in (0, \pi/4)$.
For a fully entangled initial state ($\phi \rightarrow \pi/4$) this reproduces Eq. (\ref{eq:tr_approx_av_Renyi_fullent_init_state}) using $q=2$, and for an unentangled initial state ($\phi \rightarrow 0$) one receives
\begin{equation}
\Bigl\langle H_2(\tau) \Bigr\rangle = 2 \tau + \mathcal{O}(\tau^2) \, .
\end{equation}
\section{Discussion}

In this paper we have studied how the entanglement of a two-qubit system subjected to random interactions evolves in time. We have considered two different types of randomness, namely, quenched and time-dependent random interactions, starting either from a separable or from a maximally entangled initial state. The main results are the following:
\begin{itemize}
 \item \textbf{Quenched  random interactions:} Since entanglement measures are non-linear, it makes a difference whether the average over the quenched disorder is carried out before or after evaluating the entanglement.\footnote{In general, it would be interesting to see whether there exists an inequality relating the quantities $\langle E(\rho(t)) \rangle$ and $E(\langle \rho(t)\rangle)$.}
In the first case we can compute the averaged density matrix explicitly (see Eq.~(\ref{averagedensitymatrix})), finding that the entanglement of formation decreases monotonously, see Fig.\ref{fig2}. In the second case we can only compute the so-called linear entropy~(\ref{eg:quenched_av_lin_entropy}), which is found to overshoot before it saturates.
 \item \textbf{Time-dependent random interactions:} As outlined in the Introduction, this problem is equivalent to solving a random walk on the $SU(4)$ group manifold. We find exact expressions for both the averaged density matrix and the averaged linear entropy, confirming that these quantities vary exponentially with time. For the averaged von-Neumann entropy as a special case of Tsallis entropy, however, we could only obtain a first-order approximation.
\end{itemize}

It should be noted that taking the average \textit{after} evaluating the entanglement leads to results which are not directly accessible in experiments. The reason is that the computation of entanglement in pure-state systems requires the knowledge of the full density matrix in one of the subsystems. This matrix can only be measured by means of repeated experiments under \textit{identical} conditions, which in our case would mean to use the same realization of randomness. Having estimated the entanglement this result has then to be averaged over different realizations of randomness. In experiments, where the randomness differs upon repetition, one would instead obtain a fully mixed density matrix without any information about entanglement.

As a by-product, when solving the random walk problem on the $SU(4)$ manifold we had to compute the corresponding Laplace-Beltrami operator which in turn required  to compute the metric tensor and its inverse. The computation of the metric was extremely difficult and to our best knowledge has not been done before. The resulting expressions are lengthy (see Supplemental Material). Surprisingly, we finally arrive at very simple results (see Sect.~\ref{AverageDensSection}). This indicates there might be a deeper mathematical structure behind the problem that we failed to understand so that the brute-force calculation presented here is perhaps not really necessary. It would be interesting to investigate this point in more details.

\appendix

\section{Representation of $SU(4)$ in terms of Euler angles}

The Lie algebra of the symmetry group $SU(4)$ is defined in terms of 15 Gell-Mann-like generators $\lambda_1,\ldots,\lambda_{15}$ which can be represented in various ways. In this paper we use a representation suggested in~\cite{Tilma,SSHF}. Denoting by $E_{ij}$ a $4 \times 4$ unit matrix in which the element at position $i,j$ is 1 and all others are zero, this representation is given by
\begin{equation*}
\begin{array}{lll}
\fl
\lambda_1 = E_{12}+E_{21} \quad\quad& \lambda_2 = i(E_{21}-E_{12}) \qquad\quad& \lambda_3=E_{11}-E_{22}  \\ \fl
\lambda_4 = E_{13}+E_{31} & \lambda_5 = i(E_{31}-E_{13}) & \lambda_6=E_{23}+E_{32} \\  \fl
\lambda_7 = i(E_{32}-E_{23}) & \lambda_8 = \frac{1}{\sqrt{3}}(E_{11}+E_{22}-2E_{33}) & \lambda_9=E_{14}+E_{41} \\ \fl
\lambda_{10} = i(E_{41}-E_{14}) & \lambda_{11} = E_{24}+E_{42} & \lambda_{12}=i(E_{42}-E_{24})   \\ \fl
\lambda_{13} = E_{43}+E_{34} & \lambda_{14} = i(E_{43}-E_{34}) & \lambda_{15}=\frac{1}{\sqrt{6}} (E_{11}+E_{22}+E_{33}-3E_{44}) 
\end{array}
\end{equation*}
The generators are Hermitian and obey the relations
\begin{equation}
\Tr[\lambda_i]=0\,,\qquad
\Tr[\lambda_i^2]=2\,,\qquad
[\lambda_j,\lambda_k]=2i\sum_{l=1}^4\,f_{jkl}\lambda_l\,,
\end{equation}
where $f_{jkl}$ are the $SU(4)$ structure constants defined by
\begin{equation}
f_{jkl}\;=\; \frac{1}{4i} \Tr\Bigl[ [\lambda_j,\lambda_k]\lambda_l\Bigr]\,.
\end{equation}
Group elements $U\in SU(4)$ can be generated by
\begin{eqnarray} 
U(\vect{\alpha}) &=& \; e^{i\lambda_3 \alpha_1} \; e^{i\lambda_2 \alpha_2} \; e^{i\lambda_3 \alpha_3} \; e^{i\lambda_5 \alpha_4} \; \nonumber e^{i\lambda_3 \alpha_5} \; e^{i\lambda_{10} \alpha_6} \; e^{i\lambda_3 \alpha_7} \; e^{i\lambda_2 \alpha_8} \\ && \times  \; e^{i\lambda_3 \alpha_9} \; e^{i\lambda_5 \alpha_{10}} \; e^{i\lambda_3 \alpha_{11}} \; e^{i\lambda_2 \alpha_{12}} \; e^{i\lambda_3 \alpha_{13}} \; e^{i\lambda_8 \alpha_{14}} \; e^{i\lambda_{15} \alpha_{15}}\,,
\end{eqnarray}
where $\vect{\alpha}=\{\alpha_1,\alpha_2,\ldots,\alpha_{15}\}$ is a set of 15 parameters analogous to Euler angles. Their ranges are
\begin{equation}
\label{alpharange}
 \begin{array}{rl}
\alpha_2,\alpha_4,\alpha_6,\alpha_8,\alpha_{10},\alpha_{12} &\quad\in\quad  [0,\pi/2] \\
\alpha_1,\alpha_7,\alpha_{11} &\quad\in\quad  [0,\pi]\\
\alpha_3,\alpha_5,\alpha_9,\alpha_{13} &\quad\in\quad  [0,2\pi]\\
\alpha_{14} &\quad\in\quad [0,\sqrt{3}\,\pi]\\
\alpha_{15} &\quad\in\quad [0,\sqrt{8/3}\,\pi]\,.
\end{array}
\end{equation}
According to~\cite{Tilma}, the main advantage of this parametrization in these ranges is that it provides a coverage of the group manifold without overlaps, i.e., the parametrization does not overcount the manifold. There exists also a more recent representation with similar properties which is more symmetric and transparent in the definition of the angles~\cite{Hiesmayr10}. However, the complexity of the formulas turns out to be comparable.

\section{SU(4) Haar measure}
A measure $\mu$ on a compact group $G$ is called \textit{Haar measure} if it is translation-invariant under the group itself, i.e., for any subset $S\subset G$ we have $\mu(g\circ S)=g(S)$ for all $g \in G$. Loosely speaking the Haar measure may be thought of as a constant probability density on the group manifold. In a given parametrization this requires to define a suitable invariant volume element on the group manifold. In the present case of $SU(4)$ with the parametrization defined above this volume element is given by
\begin{equation}
\d V_{SU(4)} \;=\; \mu(\vect{\alpha}) \,\,\d\alpha_1\,\d\alpha_2\cdots\d\alpha_{15} 
\end{equation}
with
\begin{eqnarray}
\label{eq:volume_element2}
\mu(\vect{\alpha})  &=&  \sin \left(2 \alpha _2\right) \sin \left(\alpha _4\right) \sin ^5\left(\alpha _6\right) \sin \left(2 \alpha _8\right) \sin^3\left(\alpha _{10}\right) \sin \left(2 \alpha _{12}\right) \\ 
&&  \times \cos ^3\left(\alpha _4\right) \cos \left(\alpha _6\right) \cos \left(\alpha _{10}\right) \,. \nonumber
\end{eqnarray}
The total group volume of $SU(4)$, first computed by Marinov~\cite{Marinov}, is then given by
\begin{equation}
V_{SU(4)} \;=\; \int \d V_{SU(4)} \;=\; \int\d\alpha_1\cdots\int\d\alpha_{15} \,\mu(\vect{\alpha}) \;=\;\frac{\sqrt{2}\pi^9}{3}\,,
\end{equation}
where the integration is carried out over the ranges specified in~(\ref{alpharange}).

\section{Derivation of the metric tensor, its inverse and Laplace-Beltrami-Operator of $SU(4)$}
\label{appendix:derive_metrix_and_laplace}

First we derive the metric $g$ of $SU(4)$ using the representation of the manifold given by~(\ref{eq:arbitrary_unitary_matrix}). Using the induced scalar product of matrices one can express the infinitesimal line element as
\begin{equation}
\label{eq:length_element_tr_UU}
ds^2 = \Tr \left[ dU dU^\dagger \right]
\end{equation}
In general the line element on a Riemann manifold is defined as
\begin{equation}
\label{eq:length_element_definition}
ds^2 = \sum_{i,j=1}^{15} g_{ij} d\alpha_i d\alpha_j
\end{equation}
Thus by equating coefficients in (\ref{eq:length_element_tr_UU}) and (\ref{eq:length_element_definition}) one obtains the matrix elements of the metric $g$. This metric with coefficients $g_{ij}$ has to be inverted to $g^{ij}$ in order to calculate the Laplace-Beltrami-Operator in (\ref{eq:definition_laplace_beltrami}). To this end one needs the square root of the determinant of the metric $\sqrt{|g|}$, but this is already given by (\ref{eq:volume_element2}). The inversion is difficult and was done manually for individual matrix elements.

\section{Integration of $R_{ijkl}(\tau) $ and $T_{ijkl\lambda\sigma\beta\epsilon}$}
\label{appendix:integrate_R_and_T}

The following table lists the appearing results of the integration of $R_{ijkl}(\tau)$ for the different cases of the indices.

\begin{table}[h]
\centering
\begin{tabular}{|m{3.3cm}|m{11cm}|}
\hline
Index propertjes & $R_{jklm}(\tau)$
\\
\hline
\hline\vspace{2mm} 
$  \neq k \neq l \neq m $ \vspace{2mm} 
& $\frac{1}{9} e^{-\tau ^2} \left(9-36 \tau ^2+42 \tau ^4-16 \tau ^6+2 \tau ^8\right)$
\\
\hline
$ (j = k \wedge l \neq m) \vee 
(j = m \wedge k \neq l) \vee 
(k = l \wedge j \neq m) \vee 
(l = m \wedge j \neq k)$ 
& $ \frac{1} {{1152}}e^{-\frac{\tau ^2}{2}}\left(1152-2304 \tau ^2+1104 \tau ^4-256 \tau ^6+25 \tau ^8-\tau ^{10}\right) $
\\
\hline
$ (j = l \wedge k \neq m) \vee 
(k = m \wedge j \neq l)$ 
& $\frac{1}{384} e^{-\frac{3 \tau ^2}{2}} \big(384-2304 \tau ^2+3312 \tau ^4-1664 \tau ^6+387 \tau ^8-27 \tau ^{10}\big) $
\\
\hline \vspace{2mm}
$ j = l\wedge k = m $ \vspace{2mm} 
& $\frac{1}{9} e^{-2 \tau ^2} \big(9-72 \tau ^2+138 \tau ^4-128 \tau ^6+50 \tau ^8-8 \tau ^{10}\big)$
\\
\hline
$ (j = k \wedge l = m) \vee 
(j = m \wedge k = l) $ 
& $ 1$
\\
\hline

\end{tabular}
\caption{Results for the integration of $R_{ijkl}(\tau)$}
\end{table}

\noindent
Averaging $T_{ijkl\lambda\sigma\beta\epsilon}$ for the initial state $\ket{\psi_0} = \ket{4} $ means to average the 4-component of the unitary matrix defined by (\ref{eq:rewrite_ev_H_as_U_i}). Thus one can rewrite $T_{ijkl\lambda\sigma\beta\epsilon}$ as
\begin{equation}
\fl\qquad\qquad
T_{ijkl\lambda\sigma\beta\epsilon} = \langle U_{i,4}^* U_{j,4} U_{k,4}^* U_{l,4} U_{i, 2 (\lambda - 1) + \sigma} U_{i, 2 (\beta - 1) + \sigma}^* U_{i, 2 (\beta - 1) + \epsilon} U_{i, 2 (\lambda - 1) + \epsilon}^* \rangle_\alphasub
\end{equation}
where we use the mapping for the qubit basis (\ref{eq:qubit_basis})
\begin{equation}
\ket{\alpha} \otimes \ket{\beta} = \ket{2 \alpha + \beta + 1}.
\end{equation}
Not all of these integrals have to be carried out.
Since the unitary matrix is randomly distributed according to Haar-measure, the rows and columns can be swapped as desired without changing the result, e.g. $T_{12141222} = T_{12411222}$.

\section{Numeric calculations}
\label{sec:numeric_calculations}
Depending on the case of the randomness we use different approaches for the numerical calculation. 
\begin{itemize}
\item 
In quenched randomness case it is possible to compute $\rho(t)$ directly for a single member of the ensemble using Eq. (\ref{eq:rho}), because $H$ is random but constant.
\item
In the temporal case we have to choose a new random Hamiltonian at each numerical time step $\d t$. The unitary operator is chosen as $U = A^{-1}\cdot A^{\dagger}$ with $A = 1 + i H \frac{\d t}{2}$ to ensure a unitary transformation. To ensure scale invariance in time, one has to scale a chosen $H$ by $1/\sqrt{\d t}$ as expected for noise terms.
\end{itemize}
Afterwards one can carry out the average over the ensemble to compute the averaged density matrix and the entanglement of formation, or to obtain the averaged entanglement measure (linear, von-Neumann).

The ensemble of the random Hamiltonians $H$ is the GUE. To obtain a $H\in $ GUE one has to chose a Hermitian matrix containing
\begin{itemize}
\item
random numbers $z$ with $z \in \mathcal{N}(0,\sigma^2)$ on the diagonal entries and
\item
complex random numbers $\frac{z_1 + i z_2}{\sqrt{2}}$  with $z_1,z_2 \in \mathcal{N}(0,\sigma^2)$ on the off-diagonal entries, where $\mathcal{N}(0,\sigma^2)$ is the normal distribution.
\end{itemize}

\section{Motivation for the diffusion equation on $SU(4)$}
\label{appendix:derive_diffusuion_equation}

In the following we would like to sketch how to derive the diffusion equation (\ref{eq:heat_equation}) starting from the time evolution operator $U(t)$ in (\ref{eq:unitary_time_evolution_operator}) in the case of temporal random interactions.
The time evolution operator $U(t)$ can be written as product of $N=\frac{t}{\d t}$ unitary transformations, that act with a new random Hamiltonian $H_i$ in the time step $i$ with constant infinitesimal time interval $\d t$:
\begin{eqnarray}
\ket{\psi(t)} &=& U(t) \ket{\psi(0)} = U_N(\d t) U_{N-1}(\d t) \cdots U_1(\d t) U_0(\d t) \ket{\psi(0)}
\nonumber\\
&=&
\left(\prod_{i=0}^{N} U_i(\d t)\right) \ket{\psi(0)} = 
\left(\prod_{i=0}^{N} e^{- i H_i \d t}\right) \ket{\psi(0)} = 
\nonumber\\
&=&
\left( \prod_{t'=0}^{t} e^{- i H_{t'} \d t} \right) \ket{\psi(0)}
\end{eqnarray}
with $U_0(dt) = 1$, so $H_0 = 0$. In this expression we rename the position index $i$ by a temporal index $t'$. Moreover, we define
\begin{equation}
\label{eq:def_dH}
\d H_{t'} = H_{t'} \d t,
\end{equation}
so that
\begin{equation}
U(t) = \prod_{t'=0}^{t} e^{- i \d H_{t'}}.
\end{equation}
In the following we want to show that the defined time evolution operator $U(t)$ equates a stochastic differential equation of a unitary Brownian motion (UBM), it the randomly chosen Hamilton $H_{t'}$ operators are drown out of
\begin{equation}
\label{eq:choose_H_from_GUE_normed_by_sqrt_dt}
H_{t'} = -Z_\GUE /{\sqrt{\d t}}
\end{equation} 
Here the $-Z_\GUE$ are randomly chosen matrices of the GUE using the normal distribution $\mathcal{N}(0,\sigma^2)$.
The sign can be chosen freely because of the symmetry around zero.
Using Eq. (\ref{eq:def_dH}) and (\ref{eq:choose_H_from_GUE_normed_by_sqrt_dt}) we get
\begin{equation}
\label{eq:def_dH_final}
\d H_{t'} = -Z_\GUE \sqrt{\d t} \;.
\end{equation}
These are random matrices according to the GUE whose components are random numbers out of $\mathcal{N}(0,\sigma^2 \d t)$.
Introducing a normalized time $t\rightarrow \frac{\tau}{\sigma^2}$ the random distribution changes to $\mathcal{N}(0,\d \tau)$, which is why $\d H_{\tau'}$ generates a Wiener process. 

Having a look at the differential $\d U(\tau)$ we conclude that
\begin{eqnarray}
\label{eq:ito_for_U(t)}
\d U(\tau) &=& U(\tau + \d \tau) - U(\tau) \nonumber \\
 &=& \left(\prod_{\tau'=0}^{\tau+\d\tau} U(t') \right) - \left(\prod_{\tau'=0}^\tau U(\tau') \right) \nonumber \\
 &=& \left(\prod_{\tau'=0}^\tau U(\tau') \right) \left(U(\tau+\d\tau) - 1 \right) \nonumber \\
 &=& U(\tau) \left(e^{-i \d H_\tau } - 1 \right) \nonumber \\
 &=&  i U(\tau) \d H_\tau- \frac{1}{2} U(\tau) \d \tau,
\end{eqnarray}
where we expanded the exponential function in the last step for $\d \tau\rightarrow 0$ using (\ref{eq:def_dH_final}). This is a stochastic differential equation for the Ito process  $\{U(\tau)\}_{\tau\in\mathbb{R}^+}$ and therefore describes a unitary Brownian motion on $SU(4)$ (see also \cite{Benaych,Levy}).

As stated by \cite{Nechita13} the Markov generator $A$ of this diffusion process is given by the Laplace-Beltrami-Operator $\triangle$ on the Riemannian manifold $A = \frac{1}{2}\triangle$.
As described above, the metric is induced by the matrix scalar product.
Since the Laplace-Beltrami-Operator is independent of the choice of the basis, we use an operator $\triangle_\alphavec$ that is parametrized by the Euler angles $\alphavec$, getting
\begin{equation}
\frac{\partial p(\alphavec,\tau)}{\partial \tau} = \frac{1}{2} \triangle_\alphavec p(\alphavec,\tau).
\end{equation}
Using the rescaled time $t = \frac{\tau}{\sigma^2}$ and the definition  $D\equiv\frac{\sigma^2}{2}$ we end up with the diffusion equation  (\ref{eq:heat_equation}).

\vspace{5mm}
\noindent
\textbf{Acknowledgments}\\
J. U. would like to thank NRF Grant No.2013R1A6A3A03028463 for financial support.


\end{document}